\documentclass[showkeys,superscriptaddress,amsmath, amssymb, aps]{revtex4}

\begin{document}
\title{Whether an enormously large energy density of the quantum vacuum
is catastrophic}
\author{Vladimir~M.~Mostepanenko}
	\affiliation{Central Astronomical Observatory at Pulkovo
		of the Russian Academy of Sciences, St.Petersburg, 196140, Russia}
	\affiliation{Institute of Physics, Nanotechnology and
		Telecommunications, Peter the Great Saint Petersburg
		Polytechnic University, St.Petersburg, 195251, Russia}
	\affiliation{Kazan Federal University, Kazan, 420008, Russia}
\author{Galina~L.~Klimchitskaya}
	\affiliation{Central Astronomical Observatory at Pulkovo
		of the Russian Academy of Sciences, St.Petersburg, 196140, Russia}
	\affiliation{Institute of Physics, Nanotechnology and
		Telecommunications, Peter the Great Saint Petersburg
		Polytechnic University, St.Petersburg, 195251, Russia}

\begin{abstract}
The problem of an enormously large energy density of
{ the} quantum
vacuum is discussed in connection with the concept of renormalization
of physical parameters in quantum field theory. Using the method of
dimensional regularization, it is recalled that the normal ordering
procedure of creation and annihilation operators is equivalent to a
renormalization of the cosmological constant leading to its zero and
nonzero values in Minkowski space-time and in the standard cosmological
model, respectively.
{It is argued that  a frequently discussed gravitational
effect, resulting from an enormously large energy density described
by the nonrenormalized (bare) cosmological constant, might be nonobservable
much like some other bare quantities introduced in the formalism of
quantum field theory}.
\end{abstract}

\maketitle

\section{Introduction}
The elaboration of quantum field theory has raised a number of fundamental
problems that remain to be finally resolved. One of them is the problem of
quantum vacuum which is of crucial importance for physics of elementary
particles and cosmology. It has been known that the vacuum stress-energy
tensor of quantized fields diverges at large momenta. In the framework of
 quantum field theory, the special procedure was elaborated which
makes the vacuum expectation values of  physical observables,
{ such as the energy density and pressure}, equal to
zero. It is the normal ordering of creation and annihilation operators
{  (note, however, that nonzero vacuum expectation values may
occur due to the spontaneous symmetry breaking)}.
This was considered as wholly reasonable as soon as all branches of physics
with the only exception of gravitation deal not with the absolute values of
energy, but with energy differences. In so doing, all energies are measured
from the infinite vacuum energy.

In succeeding years it was understood, however, that an energy density of
quantum vacuum should have incredibly large gravitational effects which are
not observed experimentally. It is common to assume that the local quantum
field theory is valid up to the Planck energy scale
$E_{\rm Pl}\sim 10^{19}~\mbox{Gev}\sim 10^9~$J. If one makes a cutoff in
the divergent vacuum stress-energy tensor at respective Planck momentum
$p_{\rm Pl}=E_{\rm Pl}/c$, the obtained energy density
$\sim 10^{111}~\mbox{J/m}^3$ would exceed the observed value
$\sim 10^{-9}~\mbox{J/m}^3$ associated with an accelerated expansion of
the Universe by the factor of $10^{120}$ \cite{1,2}.
{ Taking into account that the vacuum stress-energy tensor of quantized
fields can be described in terms of the cosmological constant \cite{2a}},
the same discrepancy is obtained between its theoretical and experimental
values giving a reason to speak about the ``vacuum catastrophe" \cite{3}.
At the moment it is widely believed that this is
{  one of the biggest  unsolved
problems of modern physics \cite{3a}}.

An assumption that the zero-point oscillations should gravitate similar
to real elementary particles could be doubted because there is no direct
way to check it out experimentally. In this paper the problem of vacuum
energy is discussed in connection with the concept of renormalizations
in quantum field theory. Using the method of dimensional regularization,
we demonstrate rigorously that the vacuum stress-energy tensor of quantized
fields is proportional to the metrical tensor. On this basis, it is argued
that the bare (nonrenormalized) value of the cosmological constant might be
excluded from theoretical description of the measurement results much as
it holds for the bare electron mass and charge in standard quantum
electrodynamics.

In Section~2, the divergences in the vacuum stress-energy tensor are
discussed in connection with other divergences in quantum field theory.
Section~3 contains the dimensional regularization of the vacuum
stress-energy tensor. In Section~4, the problem of the vacuum energy
density is considered in the context of gravitational theory.
Section~5 contains our conclusions and discussion.

The units with $\hbar=c=1$ are used throughout the paper.

\section{Divergences in quantum field theory and the quantum vacuum}

It is common knowledge that in renormalizable quantum field theories the
matrix elements of physical observables are identically expressed in
terms of bare and real (physical) quantities (for instance, bare and real
charge and mass of an electron in quantum electrodynamics).
In so doing, the real and bare quantities differ by a formally infinitely large factors. It is necessary to stress that bare (i.e., noninteracting) physical
objects are nonobservable because any observation must be accompanied by
some interaction \cite{4}. In terms of bare quantities the most of matrix
elements of physical observables are expressed by the divergent integrals and,
thus, are infinitely large. However, being expressed in terms of real
quantities, these matrix elements become finite and in excellent
agreement with the measurement data. Because of this, a removal of divergences
by means of going from the bare to real  quantities (the so-called renormalization
procedure) can be considered as quite satisfactory. As to the status of bare
 objects, they might be treated as having a little physical importance.

The vacuum stress-energy tensor of quantized fields is characterized by
a higher (fourth-order) divergence in comparison with the matrix elements
of the normally ordered operators. As discussed in Section~1, the normal
ordering procedure makes equal to zero the vacuum expectation values of
main physical observables.
{  (Note that in nonlinear quantum field theories the validity
of this statement depends on whether or not the vacuum state exists and on
the specific form of interaction.)}
According to the postulate of quantum
field theory, the operators of all physical quantities are expressed via
the operators of fields in the same way as in classical field theory,
and the normal ordering procedure establishes the proper order of operator
multiplication \cite{5}. In fact, this postulate is an application of the
correspondence principle and the normal ordering procedure makes it
unambiguous. As a result, the stress-energy tensor of the zero-point
oscillations (the so-called virtual particles) is simply disregarded.
For this reason, one may believe that the virtual particles in themselves
are not observable and, specifically, do not gravitate.

There are, however, many effects where the zero-point oscillations
contribute indirectly as a result of some interaction.
{  In perturbation theory, the pure vacuum diagrams look
like the closed loops with no external legs. They are characterized by the
fourth-order divergence and give rise to the vacuum energy density.
Many diagrams, however, have internal loops and also external legs
representing real particles. These diagrams describe experimentally
observable quantities, such as the anomalous magnetic moment of an
electron or the Lamb shift, whose values are partially determined by
the quantum vacuum.
One should mention also}
the Casimir effect where the spectrum of zero-point oscillations is altered
by an interaction with the material boundaries \cite{6,7,8}.
It is important to underline, however, that the finite Casimir energy
density and force acting between the boundary surfaces are obtained after
subtracting a divergent energy density of the quantum vacuum  in an
unrestricted space, i.e., disregarding the same quantity as in standard
quantum field theory. It cannot be too highly stressed that only this
finite and measurable energy density is a source of gravitational
interaction \cite{9,10,10a}. The Casimir-like energy density
also arises in spaces with nontrivial topology due to identification
conditions imposed on the quantized fields \cite{11,12,13}.
{  Note that there is an approach which describes  the
Casimir effect as relativistic quantum forces between charges and
currents in the material boundaries without reference to zero-point
energies \cite{13a}. It seems, however, that along these lines it
would be difficult to describe the Casimir energy density arising in
topologically nontrivial spaces in cosmology because the identification
conditions are not caused by a matter and do not involve any charges
and currents.}

The question arises whether it is possible to treat the normal ordering
procedure in terms of renormalization. Before answering it in the next
section, we briefly recall an explicit form of the vacuum stress-energy
tensor of quantized fields. For this purpose, one degree of freedom of
a quantized field can be modelled by the real scalar field
$\varphi(x)$ of mass $m$ having the Lagrangian density
\begin{equation}
L(x)=\frac{1}{2}\left[\partial_k\varphi(x)\partial^k\varphi(x)
   -m^2\varphi^2(x)\right]
\label{eq1}
\end{equation}
\noindent
and the stress-energy tensor
\begin{equation}
T_{ij}(x)=\partial_i\varphi(x)\partial_j\varphi(x)
-L(x)g_{ij}.
\label{eq2}
\end{equation}
\noindent
Here, $i,\,j,\,k=0,\,1,\,2,\,3$, and the metrical tensor is
$g_{ij}={\rm diag}(1,\,-1,\,-1,\,-1)$.

The field operator is presented in the form
\begin{equation}
\varphi(x)=\frac{1}{(2\pi)^{3/2}}\int
\frac{d^3\mbox{\boldmath$p$}}{\sqrt{2\omega(\mbox{\boldmath$p$})}}\left(
e^{-ipx}c_{\mbox{\boldmath$p$}}+
e^{ipx}c_{\mbox{\boldmath$p$}}^{+}\right),
\label{eq3}
\end{equation}
\noindent
where $\omega(\mbox{\boldmath$p$})=(m^2+ \mbox{\boldmath$p$}^2)^{1/2}$,
$px=\omega t-\mbox{\boldmath$px$}$ and
$c_{\mbox{\boldmath$p$}}$, $c_{\mbox{\boldmath$p$}}^{+}$ are the
annihilation and creation operators satisfying the standard commutation
conditions. The vacuum state is defined by
\begin{equation}
c_{\mbox{\boldmath$p$}}|0\rangle{ =0},
\qquad
\langle 0|c_{\mbox{\boldmath$p$}}^{+}=0.
\label{eq4}
\end{equation}

Substituting Equations~(\ref{eq1}) and (\ref{eq3}) in Equation~(\ref{eq2})
for $i=j=0$ and using Equation~(\ref{eq4}), one finds
\begin{equation}
\langle 0|T_{00}(x)|0\rangle=\frac{1}{2(2\pi)^3}\int
d^3\mbox{\boldmath$p$}\,\omega(\mbox{\boldmath$p$}).
\label{eq5}
\end{equation}
\noindent
In a similar way, from  Equation~(\ref{eq2}) at
$i=j=\mu=1,\,2,\,3$ we obtain the common result for all diagonal
components
\begin{equation}
\langle 0|T_{\mu\mu}(x)|0\rangle=\frac{1}{2(2\pi)^3}\int
\frac{d^3\mbox{\boldmath$p$}}{\omega(\mbox{\boldmath$p$})}
p_{\mu}^2.
\label{eq6}
\end{equation}

It is seen that the quantities (\ref{eq5}) and (\ref{eq6}) diverge as $p^4$
at high momenta. This is because the stress-energy tensor was not normally
ordered with respect to the creation and annihilation operators; otherwise
the zero results would be obtained. As to the nondiagonal components of
the vacuum stress-energy tensor, they are equal to zero,
\begin{equation}
\langle 0|T_{0\mu}(x)|0\rangle=0,
\label{eq7}
\end{equation}
\noindent
whether or not the normal ordering procedure is used.

Equations (\ref{eq5})--(\ref{eq7}) can be identically rewritten in the
form of one equation
\begin{equation}
\langle 0|T_{\mu\mu}(x)|0\rangle=\frac{1}{2(2\pi)^3}\int
d^3p\frac{p_ip_j}{\omega(\mbox{\boldmath$p$})}.
\label{eq8}
\end{equation}

Equation (\ref{eq8}) is in fact applicable to one degree of freedom of
any bosonic or fermionic field \cite{2}. In the latter case the
commutation relations for creation and annihilation operators should
be replaced with the anticommutation ones and the sign minus appears
in front of the right-hand side in Equation (\ref{eq8}).
Let we have $P$ bosonic fields $\varphi_1,\,\ldots,\,\varphi_{P}$
with masses $m_1,\,\ldots,\,m_{P}$ and $g_1,\,\ldots,\,g_{P}$ degrees
of freedom each and $Q$ fermionic fields
$\psi_1,\,\ldots,\,\psi_{Q}$
with masses $M_1,\,\ldots,\,M_{Q}$ and $h_1,\,\ldots,\,h_{Q}$ degrees
of freedom, respectively. In this case the vacuum stress-energy tensor
takes the form
\begin{equation}
\langle 0|T_{ij}^{{\rm tot}}(x)|0\rangle=\frac{1}{2(2\pi)^3}\int
d^3p\,p_ip_j\left(
\sum_{l=1}^{P}\frac{g_l}{\sqrt{m_l^2+\mbox{\boldmath$p$}^2}}-
\sum_{l=1}^{Q}\frac{h_l}{\sqrt{M_l^2+\mbox{\boldmath$p$}^2}}
\right).
\label{eq9}
\end{equation}
\noindent
In spite of the fact that this quantity contains both positive and negative contributions, it remains divergent and turns into zero only in the case
of exact supersymmetry which is not supported by the experimental data.
In the next sections we discuss the possibility to remove this infinity
by means of renormalization.

\section{Geometric structure of the vacuum stress-energy tensor}

It is a common assumption based on the relativistic covariance that the
vacuum stress-energy tensor is proportional to a constant times the
metrical tensor. In such a manner the vacuum energy density can be
identified up to a multiple with the cosmological constant which might
be a subject of renormalization similar to some other parameters in
quantum field theory. Taking into account the fundamental importance of
this issue, it seems useful from at least a pedagogical point of
view to perform an immediate analysis of the geometrical structure of
Equations (\ref{eq6})--(\ref{eq9}). This can be made by the method of
dimensional regularization \cite{14}.

We start with generalization of Equations (\ref{eq5}) and (\ref{eq6})
to the case of $N$-dimensional space-time having $N-1$ spatial dimensions.
{}From Equation (\ref{eq5})  one obtains
\begin{equation}
\langle 0|T_{00}(x)|0\rangle_N=\frac{1}{2(2\pi)^{N-1}}\int
d^{N-1}p\,\sqrt{m^2+\mbox{\boldmath$p$}^2},
\label{eq10}
\end{equation}
\noindent
where the volume element and the second power of momentum are given by
\begin{equation}
d^{N-1}p=p^{N-2}dpd\Omega_{N-2}, \qquad
\mbox{\boldmath$p$}^2=p^2=
\sum_{\mu=1}^{N-1}p_{\mu}^2
\label{eq11}
\end{equation}
\noindent
and the surface element of a unit sphere in $(N-1)$-dimensional space
is expressed in spherical coordinates as
\begin{equation}
d\Omega_{N-2}=d\varphi\sin\theta_1d\theta_1\sin^2\theta_2d\theta_2
\ldots\sin^{N-3}\theta_{N-3}d\theta_{N-3}.
\label{eq11a}
\end{equation}

Taking into consideration that all quantities (\ref{eq6}) with different
$\mu$ are equal in value due to the isotropy of space, the generalization of
Equation (\ref{eq6}) to $N$-dimensional case takes the form
\begin{equation}
\langle 0|T_{\mu\mu}(x)|0\rangle_N=\frac{1}{2(2\pi)^{N-1}(N-1)}
\int\frac{p^2d^{N-1}p}{\sqrt{m^2+p^2}}
\delta_{\mu\mu},
\label{eq12}
\end{equation}
\noindent
where $\delta_{\mu\mu}=1$ are the diagonal elements of Kronecker's symbol.

Integrating in Equations (\ref{eq10}) and (\ref{eq12}) with the help of
Equations (\ref{eq11}),  (\ref{eq11a}) and the equality
\begin{equation}
\int d\Omega_{N-2}=
\frac{2\pi^{\frac{N-1}{2}}}{\Gamma\left(\frac{N-1}{2}\right)},
\label{eq13}
\end{equation}
\noindent
where $\Gamma(z)$ is the gamma function, we arrive at
\begin{eqnarray}
&&
\langle 0|T_{00}(x)|0\rangle_N=D_N\int_{0}^{\infty}dp p^{N-2}
\sqrt{m^2+p^2},
\nonumber \\
&&
\langle 0|T_{\mu\mu}(x)|0\rangle_N=\frac{D_N}{N-1}
\int_{0}^{\infty}dp \frac{p^{N}}{\sqrt{m^2+p^2}}
\delta_{\mu\mu}.
\label{eq14}
\end{eqnarray}
\noindent
{Here the factor $D_N$ is defined as}
\begin{equation}
D_N=\frac{1}{2^{N-1}\pi^{\frac{N-1}{2}}\Gamma\left(\frac{N-1}{2}\right)}.
\label{eq15}
\end{equation}

It is convenient to introduce in Equation~(\ref{eq14}) the
dimensionless integration
variable $y=p/m$ and obtain
\begin{eqnarray}
&&
\langle 0|T_{00}(x)|0\rangle_N=D_Nm^N\int_{0}^{\infty}dy y^{N-2}
\sqrt{1+y^2},
\nonumber \\
&&
\langle 0|T_{\mu\mu}(x)|0\rangle_N=\frac{D_Nm^N}{N-1}
\int_{0}^{\infty}dy \frac{y^{N}}{\sqrt{1+y^2}}
\delta_{\mu\mu}.
\label{eq16}
\end{eqnarray}

Now we put $N=4+2\varepsilon$ where $\varepsilon$ is, in general,  a
complex parameter. In this case Equations (\ref{eq16}) are rewritten as
\begin{eqnarray}
&&
\langle 0|T_{00}(x)|0\rangle_{4+2\varepsilon}=
D_{4+2\varepsilon}m^4\left(\frac{m}{m_f}\right)^{2\varepsilon}
\int_{0}^{\infty}dy y^{2+2\varepsilon}
\sqrt{1+y^2},
\nonumber \\
&&
\langle 0|T_{\mu\mu}(x)|0\rangle_{4+2\varepsilon}=
\frac{D_{4+2\varepsilon}m^4}{3+2\varepsilon}
\left(\frac{m}{m_f}\right)^{2\varepsilon}
\int_{0}^{\infty}dy \frac{y^{4+2\varepsilon}}{\sqrt{1+y^2}}
\delta_{\mu\mu},
\label{eq17}
\end{eqnarray}
\noindent
where $m_f$ is a fictitious mass introduced in order
to make the dimension of
quantities (\ref{eq17}), written in the space-time of $4+2\varepsilon$
dimensions, the same as for $N=4$. It is significant that,
whereas the quantities (\ref{eq16}) are divergent, the quantities
(\ref{eq17}) converge in terms of the boundary values of distributions
when ${\rm Im}\varepsilon\neq 0$.

The integrals in the quantities (\ref{eq17}) can be calculated in terms
of the beta function \cite{15}
\begin{equation}
\int_{0}^{\infty}\frac{y^{2x-1}dy}{(1+y^2)^{x+t}}=
\frac{1}{2}{\rm B}(x,t)
\label{eq18}
\end{equation}
\noindent
with the result
\begin{eqnarray}
&&
\langle 0|T_{00}(x)|0\rangle_{4+2\varepsilon}=\frac{1}{2}
D_{4+2\varepsilon}m^4\left(\frac{m}{m_f}\right)^{2\varepsilon}
{\rm B}\left(\frac{3+2\varepsilon}{2},-2-\varepsilon\right),
\nonumber \\
&&
\langle 0|T_{\mu\mu}(x)|0\rangle_{4+2\varepsilon}=
\frac{D_{4+2\varepsilon}m^4}{2(3+2\varepsilon)}
\left(\frac{m}{m_f}\right)^{2\varepsilon}
{\rm B}\left(\frac{5+2\varepsilon}{2},-2-\varepsilon\right)
\delta_{\mu\mu}.
\label{eq19}
\end{eqnarray}
\noindent
Then, using the equality \cite{15}
\begin{equation}
{\rm B}(x,t)=\frac{\Gamma(x)\Gamma(t)}{\Gamma(x+t)},
\label{eq20}
\end{equation}
\noindent
we finally find
\begin{eqnarray}
&&
\langle 0|T_{00}(x)|0\rangle_{4+2\varepsilon}=
-\frac{m^4}{2^{5+2\varepsilon}\pi^{2+\varepsilon}}
\left(\frac{m}{m_f}\right)^{2\varepsilon}
\Gamma(-2-\varepsilon),
\nonumber \\
&&
\langle 0|T_{\mu\mu}(x)|0\rangle_{4+2\varepsilon}=
\frac{m^4}{2^{5+2\varepsilon}\pi^{2+\varepsilon}}
\left(\frac{m}{m_f}\right)^{2\varepsilon}
\Gamma(-2-\varepsilon)
\delta_{\mu\mu}.
\label{eq21}
\end{eqnarray}

{}From Equation~(\ref{eq21}) it is seen that
\begin{equation}
\langle 0|T_{\mu\mu}(x)|0\rangle_{4+2\varepsilon}=
-\langle 0|T_{00}(x)|0\rangle_{4+2\varepsilon}
\label{eq22}
\end{equation}
\noindent
for any $\varepsilon$ and, thus, the regularized vacuum stress-energy
tensor is really proportional to $g_{ij}$.

Using elementary properties of the gamma function, one obtains
\begin{equation}
\Gamma(-2-\varepsilon)=\frac{1}{2}\left[
-\frac{1}{\varepsilon}+\frac{3}{2}-\gamma+O(\varepsilon)\right],
\label{eq23}
\end{equation}
\noindent
where $\gamma=0.5772\ldots$ is the Euler constant.
Then, expanding the quantities (\ref{eq21}) in powers of $\varepsilon$ and
preserving only those terms which do not vanish in the limit
$\varepsilon\to 0$, we arrive at
\begin{equation}
\langle 0|T_{ij}(x)|0\rangle_{4+2\varepsilon}=
\frac{m^4}{64\pi^2}\left(\frac{1}{\varepsilon}-\frac{3-2\gamma}{2}
+\ln\frac{m^2}{4\pi m_f^2}\right)g_{ij}.
\label{eq24}
\end{equation}

Applying the result (\ref{eq24}) to each degree of freedom of $P$
bosonic and $Q$ fermionic fields, we can write
\begin{equation}
\langle 0|T_{ij}^{{\rm tot}}(x)|0\rangle_{4+2\varepsilon}=
I_{\varepsilon}\,g_{ij},
\label{eq25}
\end{equation}
\noindent
where the divergent in the limiting case $\varepsilon\to 0$
constant on the right-hand side is equal to
\begin{equation}
I_{\varepsilon}=\frac{1}{64\pi^2}\left[
\sum_{l=1}^{P}g_lm_l^4
\left(\frac{1}{\varepsilon}-\frac{3-2\gamma}{2}
+\ln\frac{m_l^2}{4\pi m_f^2}\right)-
\sum_{l=1}^{Q}h_lM_l^4
\left(\frac{1}{\varepsilon}-\frac{3-2\gamma}{2}
+\ln\frac{M_l^2}{4\pi m_f^2}\right)
\right].
\label{eq26}
\end{equation}

In flat Minkowski space-time the normal ordering procedure, thus, implies
a transition from an infinitely large in the limit $\varepsilon\to 0$
quantity (\ref{eq26}) to zero.
In the next section we consider the case of curved
space-time and show that such a transition is equivalent to zero
renormalized value of the  cosmological constant.

\section{The quantum vacuum and gravitation}

Having no quantum theory of gravity, it is reasonable  to
consider quantized fields on the background of curved space-time.
This is some kind of a semi-classical approach which is analogous to
quantum electrodynamics in external field and can be considered as
a one-loop approximation to the future quantum gravity \cite{16,17}.

It is easily seen \cite{17,18} that in the quasi-Euclidean Friedmann
Universe, which is a part of the standard cosmological model, the
main divergent term in the vacuum expectation value of the
stress-energy tensor of quantized fields has the same form as in
Minkowski space-time, i.e., is given by Equations (\ref{eq25}) and
(\ref{eq26}). Therefore, the self-consistent Einstein equations,
determining the space-time metric, can be written as
\begin{equation}
R_{ij}-\frac{1}{2}Rg_{ij}+\Lambda_{\varepsilon}^{(b)}g_{ij}=
-8\pi G\left[I_{\varepsilon}g_{ij}+\langle T_{ij}\rangle
+T_{ij}^{(m)}\right].
\label{eq27}
\end{equation}
\noindent
Here, $R_{ij}$ and $R$ are the Ricci tensor and the scalar curvature,
$\Lambda_{\varepsilon}^{(b)}$  is the bare (nonrenormalized)
cosmological constant, $G$ is the gravitational constant,
$\langle T_{ij}\rangle$ is a remaining part
of the vacuum stress-energy tensor after separating the leading
divergent term, and $T_{ij}^{(m)}$ is the stress-energy tensor of
classical background matter. Note that in curved space-time
$\langle T_{ij}\rangle$ may contain the lower-order divergent terms
as well as contributions  from the vacuum polarization and creation of
particles due to nonstationarity of the gravitational background.
{  As a result, the physical gravitational constant  differs
from the bare one by a divergent factor \cite{16,17}.}

Basing on Equation (\ref{eq27}), one can define the physical
(renormalized) cosmological constant by the equality
\begin{equation}
\Lambda^{\rm (ren)}=\Lambda_{\varepsilon}^{(b)}+
8\pi GI_{\varepsilon}.
\label{eq28}
\end{equation}
\noindent
In doing so, only $\Lambda^{\rm (ren)}$ enters the self-consistent
Equation~(\ref{eq27}). The value of $\Lambda^{\rm (ren)}$ should
be determined experimentally. Specifically, in Minkowski space-time
one should put $\Lambda^{\rm (ren)}=0$, whereas in the standard
cosmological model the value of $\Lambda^{\rm (ren)}$ was found
from an acceleration rate of the Universe expansion (see Section 1).
Under this approach, an infinitely large (bare) value of the
cosmological constant might be considered as of little
physical importance
similar to other bare parameters of quantum field theory.
{  It would be of course desirable to find the value of
$\Lambda^{\rm (ren)}$, or at least to justify its smallness,
theoretically. This is, however, presently impossible just as we
cannot calculate the value of the electric charge of an electron
but only to measure it.}

\section{Discussion}
\label{sec:5}
In the foregoing, we have adduced several arguments in favor of the
viewpoint that a divergent or enormously large
value of the stress-energy tensor
of the quantum vacuum might constitute not as serious problem as it is
often believed. In the framework of this approach, it is recalled
that the vacuum stress-energy tensor is proportional to the metrical
tensor and, thus, the normal ordering procedure, which helps to
obtain zero vacuum expectation values of physical observables
in flat space-time, is equivalent to the renormalization of the
cosmological constant. Unlike a widely believed opinion that
an enormously large vacuum energy density should produce large gravitational effect, we argue that the zero-point energy is not directly observable and,
in particular, does not gravitate.

\section{Conclusions}

To conclude, the renormalized (observable) value of the cosmological
constant and, thus, the physical vacuum energy density
should be determined experimentally.
It is zero in flat space-time and takes a nonzero value found from the
acceleration of the Universe expansion in cosmology.
As to the question about possible gravitational effects of an enormously
large energy density related to the nonrenormalized (bare) cosmological
constant, it may be considered more philological than physical.
The point is that the bare parameters in quantum field theory are not
directly observable, and the bare cosmological constant should not be an
exception to this rule.

It is anticipated that
future investigations of the nature of dark energy will shed new light on
the fundamental problem of quantum vacuum.


\end{document}